# Evidence of Interfacial Topological Superconductivity on the Topological Semimetal Tungsten Carbide Induced by Metal Deposition

W. L. Zhu[1,3*], X. Y. Hou[1*], J. Li[1,3*], Y. F. Huang[1,3], S. Zhang[1], J.B. He[1], D. Chen[1], M. D. Zhang[1,3], H. X. Yang[1,3,4,5], Z. A. Ren[1,3,4,5], J. P. Hu[1,3,4,5], L. Shan[1,2,3,4†], and G. F. Chen[1,3,4,5‡]

[1]Beijing National Laboratory for Condensed Matter Physics, Institute of Physics, Chinese Academy of Sciences, Beijing 100190, China

[2]Institute of Physical Science and Information Technology, Anhui University, Hefei 230601, China

[3]School of Physical Sciences, University of Chinese Academy of Sciences, Beijing 100190, China

[4]Collaborative Innovation Center of Quantum Matter, Beijing 100190, China

[5]Songshan Lake Materials Laboratory, Dongguan, Guangdong 523808, China

*These authors contributed equally to this work.

†‡To whom correspondence should be addressed. E-mail: lshan@iphy.ac.cn (L.S.); gfchen@iphy.ac.cn (G.F.C.)

**Interfaces between materials with different electronic ground states have become powerful platforms for creating and controlling novel quantum states of matter, in which inversion symmetry breaking and other effects at the interface may introduce additional electronic states. Among the emergent phenomena, superconductivity is of particular interest [1, 2]. In this work, by depositing metal films on a newly identified topological semimetal tungsten carbide (WC) single crystal, we have obtained interfacial topological superconductivity evidenced from soft point contact spectroscopy. This very robust phenomenon has been demonstrated for a wide range of Metal/WC interfaces, involving both non-magnetic and ferromagnetic films, and the superconducting transition temperatures is surprisingly insensitive to the magnetism of thin films, suggesting a spin-triplet pairing superconducting state. The results offer an opportunity to implement topological superconductivity using convenient thin film coating method.**

Searching for topological superconductors is a crucial step to find the expected new particles of Majorana fermions, and to generate Majorana zero modes which can be used as a qubit for fault-tolerant quantum computation [3-6]. Generally, there are several routes to realize topological superconductivity. One scheme is to build artificial topological superconductors based on hybrid structures, e.g., by utilizing the proximity effect between an s-wave superconductor and a spin-nondegenerate metal, e.g., the metallic surface of a 3D topological insulator [7-9]. The other scheme is to find intrinsic topological superconductors with spin-triplet odd-parity pairing. However, such a superconducting state has been very rare in nature with only very limited possible candidates [10]. A possible way to obtain intrinsic topological superconductors, although it is not guaranteed, is to dope carriers into a topological insulator or topological semimetal [11, 12]. Up to now, there have been some instructive attempts to identify an intrinsic topological superconductor in doped topological insulators [13-16]. However, the synthesis of superconducting samples with a particular doping level is difficult [17]. Furthermore, it is not easy to preserve the bulk band structures of the topological insulator after doping and to ensure the emergence of unconventional superconductivity prevailing over the conventional one [16]. On the other hand, some encouraging experiments have shown proximity-induced superconductivity on the surface of 3D topological insulators [18, 19], whereas their topological nature is yet to be confirmed. Recently, superconductivity has been observed at the point contacts formed between normal metal tips and some non-superconducting topological semimetals [20-23]. Most recently, tip-induced superconductivity was observed on WC, which is a new type of topological semimetal with super hardness [unpublished]. It was found that the induced superconductivity is insensitive to the tip's magnetism, giving a strong evidence of unconventional interfacial superconductivity [unpublished].

In this work, we show that the unconventional interfacial superconductivity possesses ubiquitously on the interface between WC single crystals and various metallic thin films. We deposited various metallic thin films on WC single crystals and performed conductance measurements on the films using soft point contact technique. Andreev reflection signal was observed in the point contact spectra for both non-magnetic Au and Pt films and ferromagnetic Fe, Co, Ni films, supporting the existence of interfacial superconductivity. The compatibility of the induced superconductivity with ferromagnetism suggests that the superconducting pairing is likely an unconventional triplet pairing. Thus, this study might open an avenue to realize topological superconductivity in a very simple way.

Figure 1a illustrates the geometry of our experiments. High-quality single crystals of WC were grown from Co-fluxes and metallic thin films were deposited on the WC surfaces by means of magnetron sputtering method with Au, Pt, Fe, Co, Ni targets, respectively. Considering the interface characteristics, the induced superconductivity should have a very small critical current. As a matter of fact, when a small current flows in the bulk sample, the measured voltage becomes very small, and very rapidly down to the limits of resolution of a nano-voltmeter as the temperature is reduced. The existence of superconductivity was examined alternatively by soft point contact technique. The contact was made between a small drop (about 30-50 μm in diameter) of Ag paste and the coated WC surface. Differential $dI/dV$~$V$ spectra were measured by the standard lock-in technique.

Figure 1b shows typical point contact spectra ($dI/dV$~$V$) measured on a Pt-coated WC. At lower temperatures, the especial spectral shape is difficult to be understood by other regimes but in good agreement with the expectation of Andreev reflection process occurs at a surface between a normal metal and a superconductor (N/S junction). In the framework of Blonder-Tinkham-Klapwijk theory [24-27], the symmetrical double peaks accompanied by a zero bias dip indicate a finite barrier at the N/S interface. Increase of barrier height will enhance the peaks and further depress the zero-bias conductance simultaneously. Another feature of the spectra is two conductance dips located outside the Andreev reflection peaks. This usually originates from critical current effect [28] that happens if the ballistic condition of $a \ll l$ cannot be satisfied, where $a$ is the contact radius and $l$ is the electron mean free path. If the thermal regime is involved, i.e., $a \gg l$, the Andreev reflection process could be completely replaced by the critical current effect, and thus a zero-bias conductance peak (ZBCP) with two dips outside it would be dominant in the spectra, as exemplified in Fig.1c. It can be seen that, with increasing temperature, the ZBCP decays continuously and finally fades away. Therefore, by plotting zero-bias resistance (ZBR)—the reciprocal of ZBCP as a function of temperature, we could obtain the superconducting transition and determine the transition temperature $T_c$. Figure 1d presents a series of ZBR~$T$ curves measured at different magnetic fields. The superconducting transition shifts towards lower temperatures with increasing fields, from which we can determine the critical field of $\mu_0 H_{c2}(T)$ as shown in Fig.1e with a zero-field transition temperature $T_c \approx 8.5$ K. As an approximation, the $\mu_0 H_{c2}(T)$ obtained in this work can be described by the well-known empirical formula of $\mu_0 H_{c2}(T) = \mu_0 H_{c2}(0)[1-(T/T_c)^2]$, which is usually appropriated for type-II superconductors.

The observation of superconductivity in such a hybrid structure is surprising by considering it is built from a normal metal and WC which is not superconducting even under a high pressure up to 11Gpa [private communications]. We have tried to make soft point contact directly on the pristine WC single crystals but cannot see any obvious superconducting signals. In combination with the results of hard point contact experiments, we could conclude that the coupling between normal metals and WC should be a key ingredient for the superconductivity observed here. As aforementioned, at low temperatures, the resistance drops rapidly below the resolution of voltmeter if we apply a small current using a standard four-point probe setup. On the contrary, it has been found that the resistance of the sample and contacts was dropped up to 72% using a modified four-point probe configuration (Supplementary Information). Therefore, we explore superconductivity using a soft point contact, which can avoid any form of surface damage. Since point contact is a micro-region probe, a statistical survey of the coated WC is necessary to confirm the existence of interfacial superconductivity. For this purpose, we have prepared more than 100 soft point contacts and usually 5 contacts on a single sample. Superconducting signal has been detected at almost all these point contacts (see Supplementary Information and refer to Fig.2a), demonstrating that this is not a local effect.

In our previous experiments of hard point contact by means of needle-anvil method, tip-induced superconductivity has been realized on WC using both non-magnetic and ferromagnetic tips. So it's meaningful to see if superconductivity could occur at the interface between WC and a ferromagnetic thin film. Figure 2a shows the spectra taken at 5 point contacts prepared on the same Co-coated WC single crystal. Either Andreev reflection signal or critical current effect can be seen in all the curves, indicating superconductivity existing everywhere. As shown in Figs.2b and 2c, temperature and magnetic field dependencies of the Andreev reflection spectra were measured in detail for a selected point contact. It can be seen the achieved superconducting state is depressed gradually with both temperature and field, exhibiting the similar behaviors as that of the Au-coated WC. Such measurements have been carried out for Fe-, Co-, and Ni-coated WC and the results are presented in Fig.3a-c, respectively. The corresponding superconducting transitions at various magnetic fields are shown in Fig.3d-f, and the determined $\mu_0 H_{c2}(T)$s are given in Figs.3g-i. It is interesting to note that the $\mu_0 H_{c2}(T)$ relations can also be described by the above mentioned empirical formula.

After a large number of repeated experiments, we can give a statistic chart of $T_c$ versus film materials in Fig.4a. Each point was obtained by measuring a zero-field ZBR~T curve or taking a series of temperature dependent point contact spectra to determine the specific temperature at which spectral features disappear completely. In this chart, it is hard to say the achieved maximum $T_c$ value has an obvious dependence on the used film materials. Most impressively, there is no direct relationship between $T_c$ and the magnetism of the materials. For the sake of further understanding, we have summarized $\mu_0 H_{c2}(0)$ versus $T_c$ as shown in Fig.4b, which was obtained by extrapolating $\mu_0 H_{c2}(T)$ curve to zero temperature. A universal trend can be seen clearly, showing a positive correlation between $\mu_0 H_{c2}(0)$ and $T_c$. Again, there is little difference between magnetic and non-magnetic films.

It is well known that ferromagnetism competes with conventional superconductivity [29]. Thus the coexistence of superconductivity and ferromagnetism is of particular interest not only for potential applications but also for the research of fundamental physics such as searching for unconventional superconductors and Majorana fermions. The interfacial superconductivity achieved at the metal film-coated WC has been proved to be insensitive to ferromagnetism, which is compatible to a spin triplet pairing state. Moreover, WC is a topological semimetal with both super hardness and high chemical stability. Using metal deposition instead of hard point contact to realize superconductivity on WC could further exclude the tip pressure or confinement effect as the dominant regime of such interfacial superconductivity. Then the coupling between normal metals and WC in some form is the most important ingredient of the superconductivity observed here. More experimental and theoretical work is needed to address this issue. Nonetheless, the realization of superconductivity on the basis of WC's topological surface states has provided a platform to explore topological superconductivity in a simple and practical way.

# Methods

## Sample preparation and characterization.

High-quality single crystals of WC were grown from Co-fluxes. Stoichiometric amounts of W and C with moderate Co were put into graphite crucible, heated to 1,700 °C, and then cooled slowly to 1,400 °C in an argon atmosphere. The residual Co-fluxes were removed by dissolving in a warm hydrochloric acid solution. The obtained single crystals are in the form of equilateral triangles with sides of 1-3 mm and thickness of 0.1-0.3 mm, and were characterized by x-ray diffraction (XRD) on a PANalytical diffractometer with Cu K$\alpha$ radiation at room temperature. The electrical transport and magnetic characters were measured using a Quantum Design PPMS-9T and a MPMS-7T SQUID VSM system, respectively.

## Sample Calibration.

Metallic thin films on WC single crystals were prepared using a sputtering deposition method with Au, Pt, Fe, Co, Ni targets, respectively. Scanning electron microscopy (Hitachi SU5000) is used to study the surface morphology of the prepared films. The film thickness was determined at a film edge using a stylus type profilometer (KLA-Tencor Profiler).

## Point-contact Spectroscopy.

Superconductivity was measured by soft point contact technique. The contact is made between a small drop (about 30-50 μm in diameter) of Ag paste and the metal film coating WC surface. The Ag electrode is connected to current and voltage leads through a thin Pt wire (18 μm in diameter) stretched over the sample. Differential d$I$/d$V$ spectra are measured by the standard lock-in technique. The DC current across the point contact is generated by current source (Keithley 6221) to bias the junction voltage, which is measured by a digital nano-voltmeter (Keithley 2182). A small ac current is generated by a lock-in amplifier (NF LI5640) after a voltage-to-current conversion. The first harmonic response of the lock-in is proportional to the differential change in the voltage d$V$.

## Acknowledgements

We acknowledge S. H. Pan for valuable discussion. This work was supported by the Ministry of Science and Technology of China (2017YFA0302904, 2018YFA0305602, 2016YFA0401000, 2016YFA0300604, 2015CB921303), the National Natural Science Foundation of China (11574372, 11322432, 11704403, 11874417) and the Chinese Academy of Sciences (XDB07020300, XDB07020100).

## Author contributions

W. L. Z. grew the single crystals with the help of J. B. H. and D. C.; W. L. Z. and Y. F. H. prepared thin films with the help of S. Z., Z. A. R.; W. L. Z., J. L., X. Y. H., Y. F. H. carried out the soft point contact spectroscopy measurements with the help of M. D. Z.; H. X. Y., W. L. Z. and M. D. Z. performed the SEM analysis; J. P. H., X. Y. H., L. S. and G. F. C. analyzed the data and wrote the manuscript. All the authors read and commented on the manuscript.

**Figure 1 Evidence of superconductivity for non-magnetic films coated WC single crystal detected by soft point contact spectra.** **a.** Schematic diagram showing the soft point-contact on the coating surface of WC sample and the differential conductance measurement electrodes. **b.** Temperature dependence of the normalized spectra in the intermediate regime for point-contact on Pt coated WC, showing Andreev reflection peaks and dips due to critical current effect. **c.** Another representative point-contact spectra obtained on Au coated WC in the thermal limit, showing a zero bias conductance peak (ZBCP) along with two dips. **d.** Magnetic field dependence of point-contact zero bias resistance showing the suppression of superconducting transition. **e.** The $\mu_0 H_{c2}$–$T_c$ phase diagram extracted from d. Each critical temperature was determined by the derivation point of the zero bias resistance curves. The red curve is fitting to the data using the empirical formula $\mu_0 H_{c2}(T) = \mu_0 H_{c2}(0)[1-(T/T_c)^2]$.

**Figure 2 d$I$/d$V$ spectra for magnetic Co-coated WC.** **a.** Andreev reflection dominated and critical current effect dominated point contact spectra acquired on different junctions on one sample coated by Co film. **b.** Temperature dependence of the normalized spectra in the intermediate regime for the junction $J_3$ in a, showing Andreev reflection peaks and dips due to critical current effect. **c.** Field dependence of the normalized spectra in b.

**Figure 3 Temperature and magnetic field evolution of the induced superconductivity. a-c.** Temperature dependence of normalized d$I$/d$V$ spectra at PCs on WC coated by magnetic Fe, Co and Ni thin films without an external magnetic field. The temperature ranges from 1.8K to 10K. **d-f.** Temperature dependence of the zero-bias resistance under several magnetic fields. **g-i.** The $\mu_0H_{c2}$–$T_c$ phase diagram extracted from d-f and fits to the data using the expression $\mu_0H_{c2}(T) = \mu_0H_{c2}(0)[1-(T/T_c)^2]$.

**Figure 4 Critical temperatures and upper critical fields. a.** Critical temperatures ($T_c$s) of the interfacial superconductivity realized on WC with diverse metal deposition. $T_c$ is determined by the onset temperature of superconducting transition from the zero bias resistance curves. **b.** Critical temperatures ($T_c$s) and upper critical fields ($\mu_0H_{c2}(0)$) of the interfacial superconductivity realized on WC with diverse metal deposition. $\mu_0H_{c2}(0)$ is determined by empirical formula $\mu_0H_{c2}(T) = \mu_0H_{c2}(0)[1-(T/T_c)^2]$ from the magnetic filed dependent ZBR. The orange dash line is a guide to the eye.

**Fig. 1**

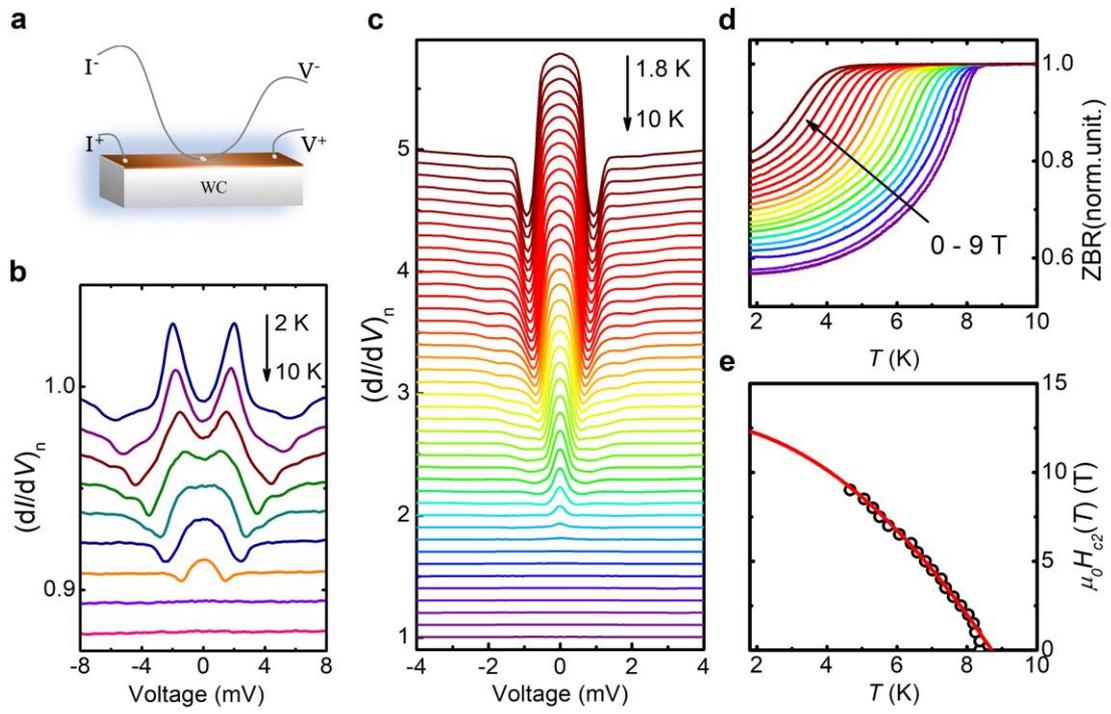

**Fig. 2**

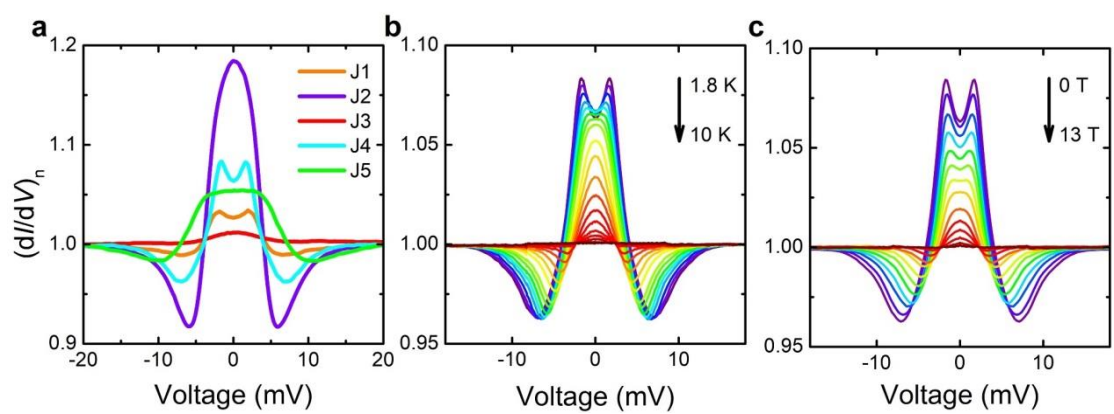

**Fig. 3**

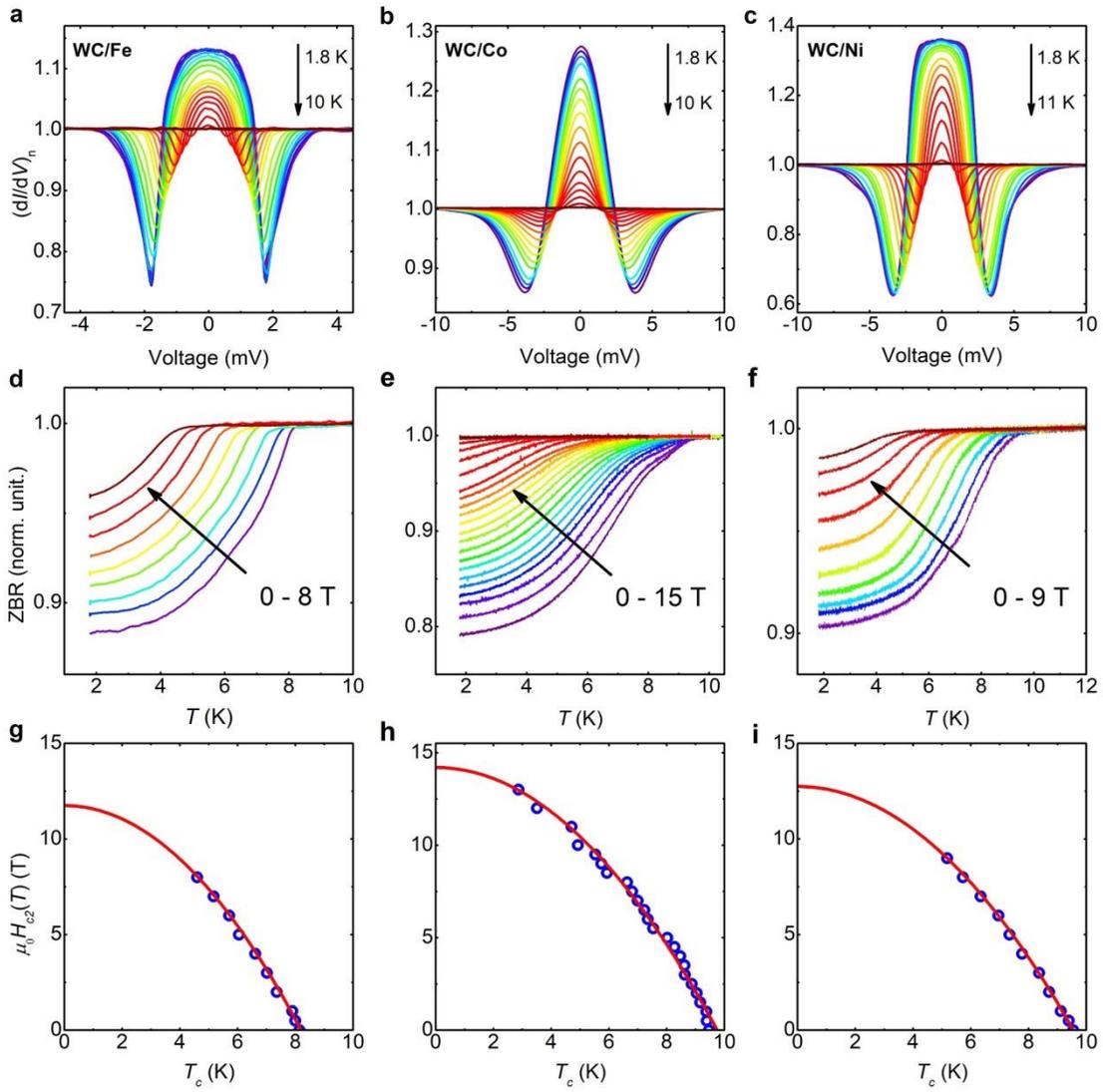

**Fig. 4**

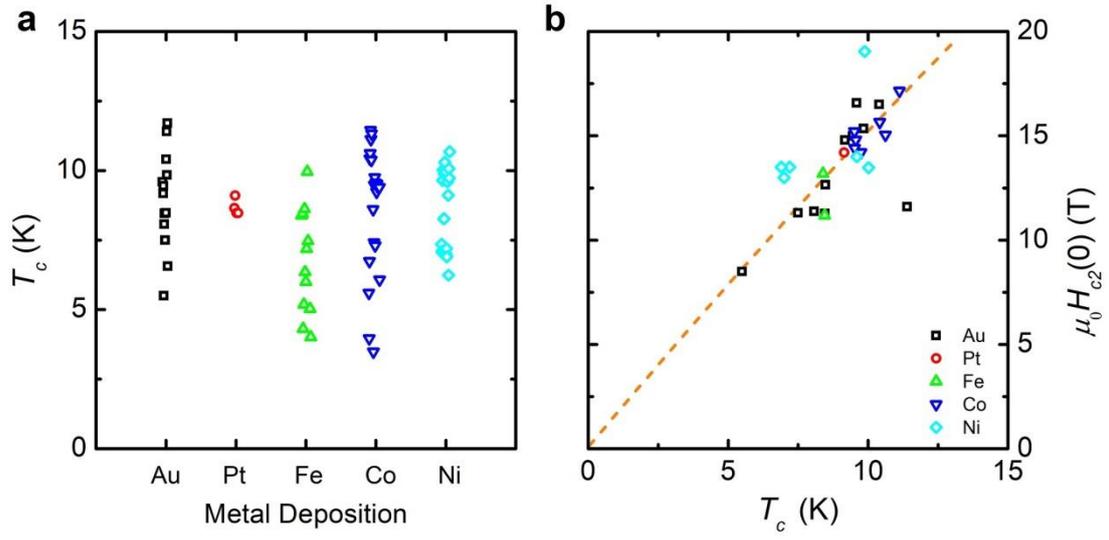

Supplementary Information for

# Evidence of Interfacial Topological Superconductivity on the Topological Semimetal Tungsten Carbide Induced by Metal Deposition

W. L. Zhu[1,3*], X. Y. Hou[1*], J. Li[1,3*], Y. F. Huang[1,3], S. Zhang[1], M. D. Zhang[1,3], H. X. Yang[1,3,4,5], Z. A. Ren[1,3,4,5], J. P. Hu[1,3,4,5], L. Shan[1,2,3,4†], and G. F. Chen[1,3,4,5‡]

[1]Beijing National Laboratory for Condensed Matter Physics, Institute of Physics, Chinese Academy of Sciences, Beijing 100190, China

[2]Institute of Physical Science and Information Technology, Anhui University, Hefei 230601, China

[3]School of Physical Sciences, University of Chinese Academy of Sciences, Beijing 100190, China and

[4]Collaborative Innovation Center of Quantum Matter, Beijing 100190, China

[5]Songshan Lake Materials Laboratory, Dongguan, Guangdong 523808, China

*These authors contributed equally to this work.

†‡To whom correspondence should be addressed. E-mail: lshan@iphy.ac.cn, (L.S.); gfchen@iphy.ac.cn, (G.F.C.)

**Soft point contact:**

In the so-called 'soft' point-contact technique, the contact is usually made on the clean sample surface by a tiny drop of Ag paste, connecting to the thin Pt wires stretched over the sample. This scale is often far larger than the electronic mean free path, however the real electrical contact occurs through some parallel nanometric channels connecting the sample surface with the individual Ag grains in a much smaller size, such as 2–10 μm. Hence these contacts can often provide spectroscopic information. This technique does not involve any pressure applied to the sample with respect to the needle–anvil technique [1]. The soft point contacts are prepared at room temperature and at ambient atmosphere.

**Transport regime:**

The interface transport regime of a point contact depends on the contact size and the electron mean free path $l$ and superconducting coherence length $\xi$ (for N/S junctions). In the ballistic regime the electron mean free path $l$ is much larger than the contact radius $a$ ($l \gg a$), and in the thermal regime on the contrary ($l \ll a$). Between these two extreme regimes, the resistance of the contact can be expressed by Wexler's formula:

$$R = \frac{2h}{e^2 a^2 k_{F,min}^2 \tau} + \Gamma(l/a)\frac{\rho_1 + \rho_2}{4a}$$

where, $\Gamma(l/a)$ is a numerical factor close to unity, $a$ is the contact diameter. $k_{F,min}$ is the minimum magnitude of the Fermi wave vector in both sides. $\rho(T)$ is resistivity of the materials.

Andreev reflection dominates the spectra characters in the ballistic regime, which can provide spectroscopic information of superconductors [2]. In the thermal regime, the resistance of the junction depends on the latter item in aforementioned formula completely. Therefore, when the superconducting transition arises in one side of N/S junction, a transition should occur on the RT curves with a tiny current flow through the junction simultaneous [3-6].

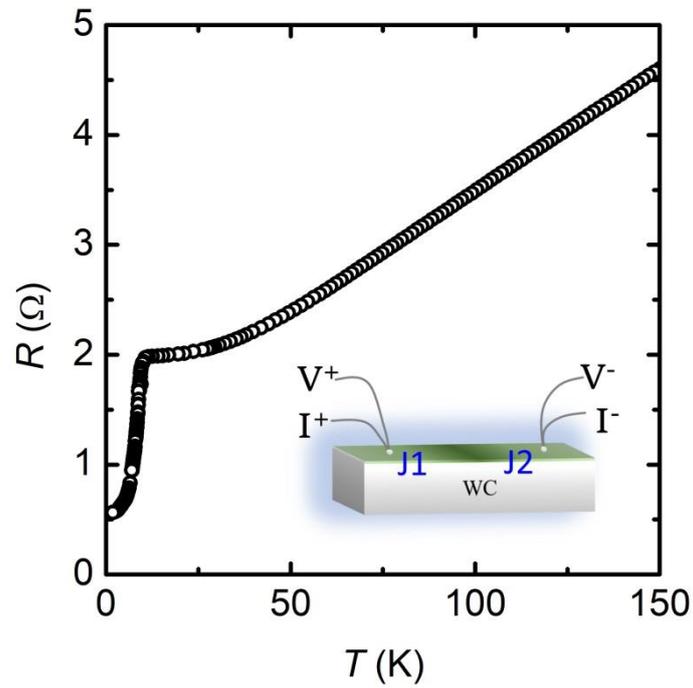

**Figure S1** Temperature dependence of the sample and junctions' resistance for a Co coated WC using a modified four-point probe setup. Below $T_c$ ∼10 K, the resistance decreases quickly to 72% of the normal value. This distinct drop mainly contributed by the N/S contact resistance in thermal limit. Inset: schematics of the modified four-probe measurement configuration.

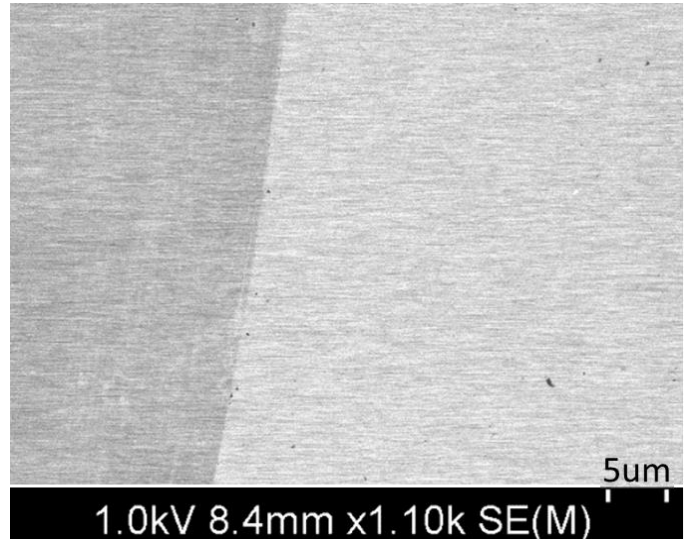

**Figure S2** High-resolution scanning electron microscopy (SEM) image of the deposited Co film on WC single crystal. The bright region is the uniform Co film. The dark region is the exposed WC surface scratched by a tweezer.

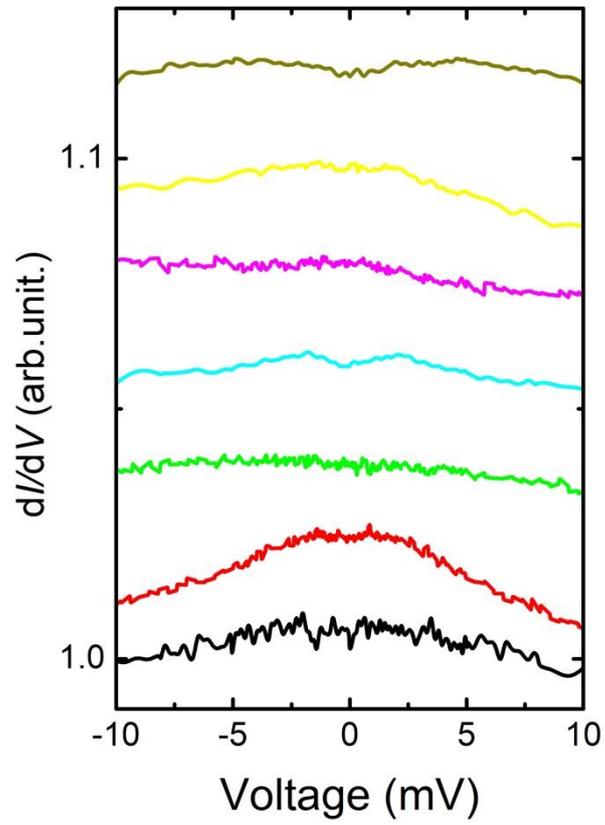

**Figure S3** Soft point contact measurements on a naked WC surface. No prominent superconducting features are observed on the spectra.

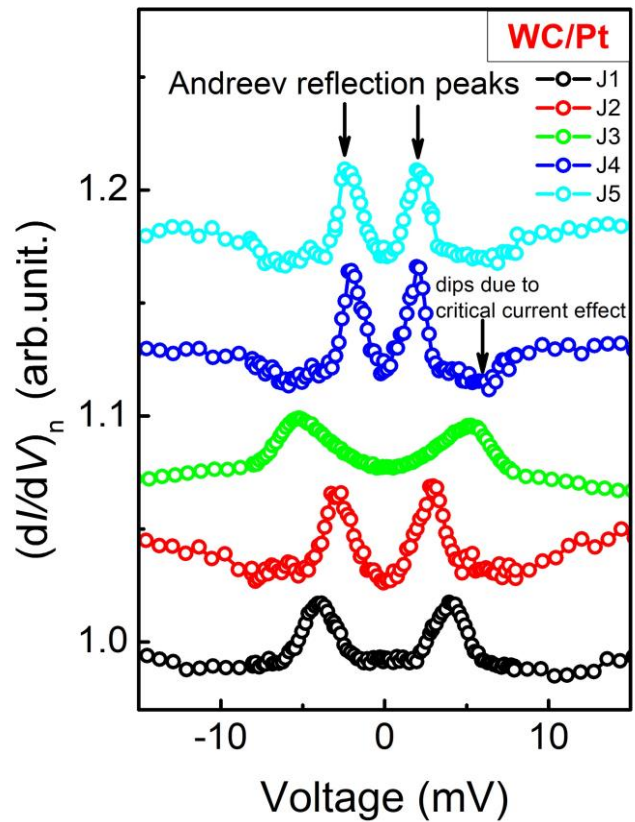

**Figure S4** Andreev reflection dominated point contact spectra acquired on different junctions on a Pt-coated WC sample.